\documentclass{ws-p8-50x6-00nofoot}
\usepackage{xspace}
\newcommand{\qsq    }{\ensuremath{Q^{2}}\xspace}
\newcommand{\qzm    }{\ensuremath{\langle \qsq \rangle}\xspace}
\newcommand{\psq    }{\ensuremath{P^{2}}\xspace}
\newcommand{\wsq    }{\ensuremath{W^{2}}\xspace}
\newcommand{\ft     }{\ensuremath{F_{2}^{\gamma}}\xspace} 
\newcommand{\ftc    }{\ensuremath{F_{2,\mathrm{c}}^{\gamma}}\xspace}
\newcommand{\ds     }{\ensuremath{D^{\star}}\xspace}
\newcommand{\eg     }{\ensuremath{E_\gamma}\xspace}
\newcommand{\xg     }{\ensuremath{x_\gamma}\xspace}
\newcommand{\etjet  }{\ensuremath{E_\mathrm{T,jet}}\xspace}
\newcommand{\etjetq }{\ensuremath{E^2_\mathrm{T,jet}}\xspace}
\newcommand{\etjeti }{\ensuremath{E_\mathrm{T,jeti}}\xspace}
\newcommand{\etajet }{\ensuremath{\eta_\mathrm{jet}}\xspace}
\newcommand{\etajeti}{\ensuremath{\eta_\mathrm{jeti}}\xspace}

\newcommand{\gevsq  }{\ensuremath{\mathrm{GeV^2}}\xspace}

\newcommand{\xvis   }{\ensuremath{x_{\mathrm{vis}}}\xspace}
\newcommand{\als    }{\ensuremath{\alpha_{s}}\xspace}
\newcommand{\epem   }{\ensuremath{{\mathrm{e}^+\mathrm{e}^-}}\xspace}
\newcommand{\jpsi   }{\ensuremath{\mathrm{J}/\psi}\xspace}
\begin{document}
\title{Summary of the structure function session at PHOTON 
       2001\footnote{Invited talk at the International Conference on
       the Structure and Interactions of the Photon, including 
       the 14th International Workshop on Photon-Photon Collisions,
       Ascona, Switzerland, 2-7 September, 2001.}
}
\author{Richard Nisius}
\address{CERN, Switzerland, E-mail: Richard.Nisius@cern.ch}
\maketitle
\abstracts{The status of and ongoing developments in the 
           measurements and theoretical studies of the structure of the photon
           have been presented at the PHOTON 2001 conference in Ascona.
           The results presented in the structure function session are 
           briefly summarised.
          }
%
%
\section{Introduction}
\label{sec:intro}
 In recent years considerable progress has been made in several
 aspects of the investigation of the photon structure.
 A recent review of the status of this field of research can be found
 in\cite{bib:NIS-9904}.
 In this paper, the most important results presented at Ascona are 
 briefly reviewed.
 Of course the material presented is intended only as an overview,
 and reflects my personal view.
 For further information of the important details of the investigations
 the reader is referred to the write-ups of the individual presentations
 elsewhere in these proceedings.
%
%
\section{Experimental Results}
\label{sec:exper}
\begin{figure}[htb]
\begin{center}
{\includegraphics[width=0.43\linewidth]{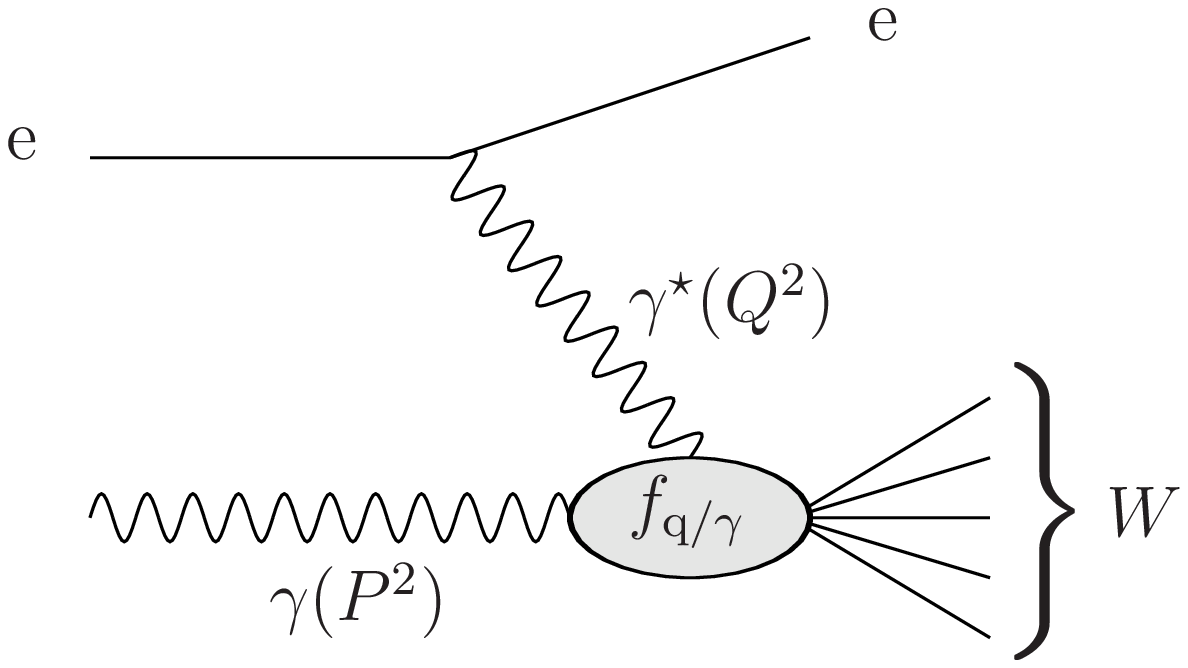}}
{\includegraphics[width=0.56\linewidth]{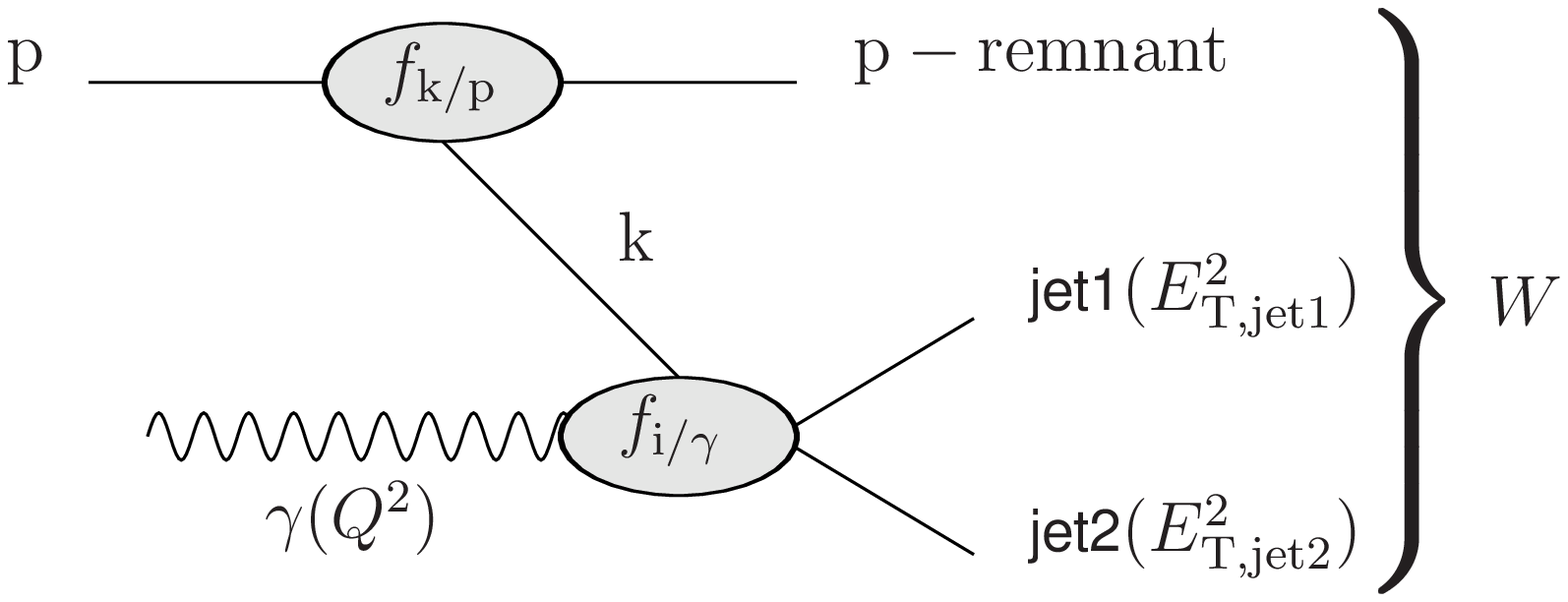}}
\caption{
         Schematic view of the reactions used in electron-photon (left) and 
         proton-photon scattering (right) for investigations of the 
         photon structure.
        }\label{fig:fig01}
\end{center}
\end{figure}

 At this conference new results on the photon structure have been presented
 by the LEP experiments, using structure functions, and by the H1 and ZEUS
 experiments, using jet cross-sections. 
 Given the different observables used in electron-photon scattering at
 LEP and proton-photon scattering at HERA, the variables used in these
 analyses are different, as illustrated in Figure~\ref{fig:fig01}.
%
%
 The factorisation scale at which the partonic structure of the photon 
 is probed is taken to be \qsq in electron-photon scattering, and
 in proton-photon scattering it is identified with the square of the 
 transverse energy of the jets \etjetq.
 The momentum fraction of the parton in the photon that takes part in the 
 hard interaction is called $x$ in electron-photon scattering, and calculated 
 from $x=\qsq/(\qsq+\wsq)$, where the quantities are shown 
 in Figure~\ref{fig:fig01}.
 In contrast, in di-jet production in proton-photon scattering this quantity
 is denoted by \xg and calculated e.g.~from the pseudorapidity \etajet and 
 transverse energy of the two jets via 
 $\xg=\sum_{\rm i} (\etjeti{\rm e}^{-\etajeti}) / (2yE)$.
 In addition, some quantities carry different names at LEP and HERA.
 The virtuality squared of the target photon $\gamma$ in
 Figure~\ref{fig:fig01}, is called \psq at LEP and \qsq at HERA,
 and the ratio of the energies of the target photon and 
 the incoming electron, $\eg/E$, is called $z$ at LEP and $y$ at HERA.
%
%
\subsection{Results from electron-photon scattering}
\label{sec:LEP}
\begin{figure}[htb]
\begin{center}
{\includegraphics[width=0.85\linewidth]{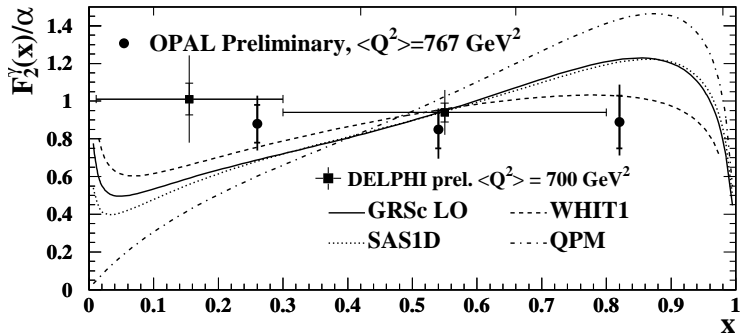}}
\caption{
         Measurement of the hadronic structure function \ft at the largest
         \qzm compared to several parametrisations.
        }\label{fig:fig02}
\end{center}
\end{figure}

 New results on the hadronic photon structure function \ft have been 
 presented both by DELPHI\cite{bib:Tiapkin} and by OPAL\cite{bib:Taylor}.
 The results are based on most of the statistics available at LEP1
 and LEP2 energies and extend the measurements of \ft to 
 $\qzm\approx 750$~\gevsq, see Figure~\ref{fig:fig02} for the highest
 values of \qzm probed to date.
 \par
%
%
 The strategy to obtain \ft from a measurement of the differential 
 cross-section is still a matter of research.
 The new DELPHI analysis uses a multivariable fit of observed distributions
 to adjust the individual components of the hadronic structure of the 
 photon when obtaining \ft. 
 This method results in larger uncertainties on \ft than what is achieved by
 using a one-dimensional unfolding of the distribution of the visible
 $x$ value of the event, \xvis, based e.g.~on the RUN 
 program\cite{bib:Blobel}, but the claim by DELPHI is that the errors
 are more reliable for the multivariable fit\cite{bib:Tiapkin}.
 \par
\begin{figure}[htb]
\begin{center}
{\includegraphics[width=0.85\linewidth]{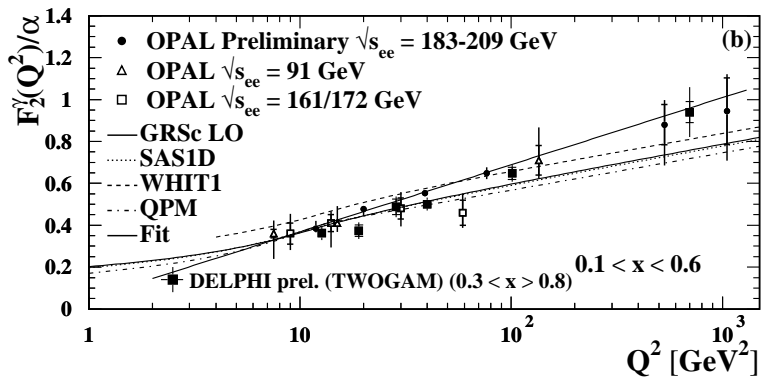}}
\caption{
         The evolution of \ft with \qsq at medium values of $x$
         compared to several parametrisations.
        }\label{fig:fig03}
\end{center}
\end{figure}

 Also the investigation of the evolution of \ft with \qsq in ranges of $x$ 
 has been continued using the LEP2 data.
 As shown in Figure~\ref{fig:fig03}, the \qsq region has been extended
 and the errors have been reduced considerably by using the 
 large luminosity available at LEP2 energies.
%
%
 With the present level of statistical precision the data start to
 challenge the existing parametrisations of \ft.
 Given this, several theoretical as well as experimental issues have to
 be addressed in more detail.
 Examples are the suppression of \ft with the virtuality squared of the 
 target photon \psq and radiative corrections to the deep-inelastic
 scattering process.
 \par
 An update has been presented\cite{bib:Csilling} of the OPAL measurement of
 the charm component \ftc using \ds mesons to identify charm quarks.
 The analysis is based on improved Monte Carlo models and higher statistics 
 compared to the published result\cite{OPALPR294}. 
 This led to an improved precision of the measurement.
 In a similar way to the structure function for light quarks, \ftc receives 
 contributions from the point-like and the hadron-like components of the 
 photon structure, as explained e.g.~in\cite{bib:NIS-9904}. 
 These two contributions are predicted\cite{bib:Laenen} to have different 
 dependences on $x$, with the hadron-like component dominating at very 
 low values of $x$ and the point-like part accounting for most of \ftc at 
 $x>0.1$.
 For $x>0.1$ the OPAL measurement is described by perturbative QCD 
 at next-to-leading order.
 For $x<0.1$ the measurement is poorly described by the NLO prediction
 using the point-like component alone, and therefore the measurement
 suggests a non-zero hadron-like component of \ftc.
 Increased statistics and a better understanding of the dynamics 
 for $x<0.1$ are needed to get a more precise measurement in this region.
 To increase the statistics it would be advantageous to combine the data 
 from the four LEP experiments.
%
%
\subsection{Results from proton-photon scattering}
\label{sec:HERA}
 New results from H1 and ZEUS have been presented concerning the
 structure of quasi-real photons\cite{bib:Valkarova}, and also of 
 virtual photons without\cite{bib:Sedlak} and with\cite{bib:West} 
 identified charm quarks.
 \par
\begin{figure}[htb]
\begin{center}
{\includegraphics[width=0.67\linewidth]{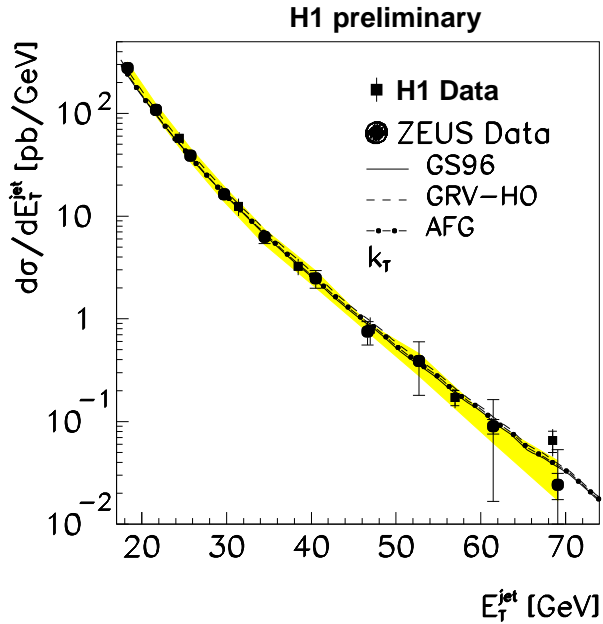}}
\caption{
         The inclusive jet cross-section as a function of \etjet
         compared to NLO predictions.
        }\label{fig:fig04}
\end{center}
\end{figure}

 There is good agreement between the ZEUS and H1 results on
 the inclusive production of jets as can be seen from Figure~\ref{fig:fig04}.
%
%
 The observed inclusive jet cross-sections are well described by existing
 parton distribution functions of the photon that have been obtained 
 from measurements of \ft.
 \par
 In contrast, there is a longstanding difference between H1 and ZEUS 
 results for di-jet final states\cite{bib:NIS-9904,bib:NIS-0101}.
 The new preliminary H1 result\cite{bib:Valkarova}, is consistent with the 
 predictions based on existing parametrisations of \ft and at present the 
 data are not precise enough to distinguish between different parametrisations.
 This has to be confronted with the earlier result from 
 ZEUS\cite{bib:ZEUS,bib:Valkarova}
 which suggested that the parton distribution functions of the photon, 
 obtained from fits to measurements of \ft made at \epem colliders, are too
 low for medium values of Bjorken $x$ and at factorisation scales of
 several hundred \gevsq.
 There are several differences between the ZEUS and H1 analyses such as the 
 choice made for the value of \als, the parton distributions used for 
 the photon, and most notably the corrections applied to the data. 
 The H1 data are corrected for detector as well as hadronisation effects and
 are shown at the partonic level. In contrast, the ZEUS results are
 corrected only for detector effects and phase space regions are selected,
 where the hadronisation corrections, as implemented in Monte Carlo models,
 are found to be small.
 It remains to be seen how much of the apparent differences between the
 results can be explained by the different analysis methods.
 Important information now also comes from LEP, where the
 parametrisations of \ft are found to be consistent with the measurements
 for factorisation scales up to 750~\gevsq, leaving less room for changes
 to the parton distribution functions of the photon.
 \par
 It is certainly desirable to complement the measurements of \ft
 with the jet measurements from HERA, which extend to even larger 
 factorisation scales, when fits for the parton distribution
 functions of the photon are performed. 
 However, first it has to be seen if a consistent picture 
 of the various HERA results can be established.
 \par
\begin{figure}[htb]
\begin{center}
{\includegraphics[width=0.86\linewidth]{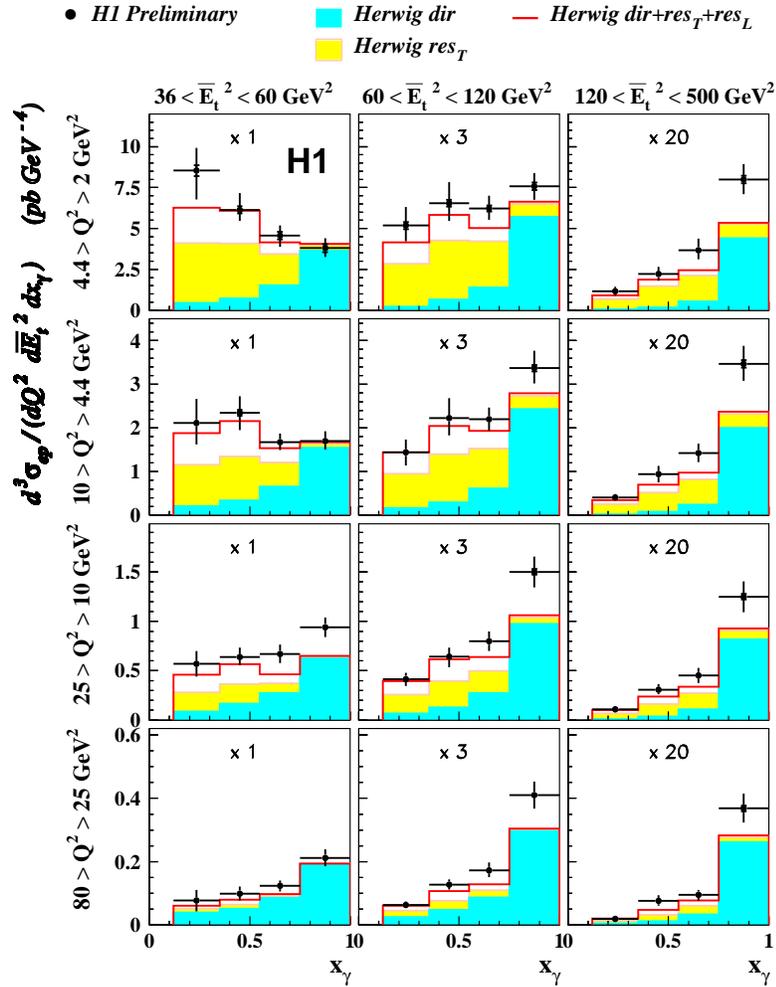}}
\caption{
         The structure of virtual photons from H1
         compared to several predictions from HERWIG.
        }\label{fig:fig05}
\end{center}
\end{figure}

 The measurement of di-jet production has been extended to the 
 investigation of the structure of virtual photons.
 In the recent ZEUS measurement\cite{bib:Sedlak} the suppression of the
 photon structure with the photon virtuality has been measured based on the
 ratio of the cross-sections for low and high values of \xg.
 The LO predictions fail to describe this ratio when using only 
 transverse virtual photons together with the SaS1D\cite{bib:SaS} 
 parametrisations of the photon structure.
 A similar difference has been found by H1. 
 In addition, it has been demonstrated by H1 that the inclusion of 
 longitudinal virtual photons helps to improve on the description of the
 observed triple differential cross-section shown in Figure~\ref{fig:fig05}.
%
%
 But the $y$ dependence of the cross-section is still not adequately
 described\cite{bib:Sedlak} for $y<0.3$. 
 More experimental as well as theoretical investigations are needed to better 
 understand these findings.
 \par
 The structure of virtual photons has also been investigated for 
 the charm component alone, using \ds mesons to identify charm quarks.
 The mass of the charm quark enters as yet another scale in the process, 
 in addition to \etjetq and \qsq.
 By again using the ratio of the cross-sections for low and high
 values of \xg it is found that the suppression with \qsq is much weaker
 in the presence of charm than for the sample containing all 
 flavours\cite{bib:West}.
 The result is less precise than the ratio for all flavours, but, in this
 case the cross-section ratio is well described by the SaS1D prediction.
%
%
\section{Theoretical Developments}
\label{sec:theo}
 A new parametrisation of \ft has been presented\cite{bib:Jankowski}.
 The extraction of \ft closely follows the procedure used by GRV\cite{bib:GRV} 
 in 1992, see also\cite{bib:NIS-9904}, with some changes and additions.
 The fit uses all modern data on \ft.
 The hadron-like input at the starting scale of the evolution based on VDM
 is no longer derived from the pion, but fitted to the data, 
 the starting scale of the evolution is varied,
 heavy quarks are included via the Bethe-Heitler process at all values of \qsq,
 and a \jpsi component is added to the sum of vector mesons.
 With these ingredients an improved fit to the data is 
 achieved\cite{bib:Jankowski} compared to the original GRV 
 parameterisation.
%
%
\section{Conclusion}
\label{sec:concl}
 The investigation of the structure of the photon is a very active field
 of research experimentally at LEP as well as at HERA, and also
 theoretically.
 Given the large statistics available at LEP2 energies and at HERA,
 the region of phase space covered is constantly increasing.
 Despite the large luminosities available, for some of the measurements
 the results are still limited by statistics and a combination of the results
 from several experiments is desirable.
 At LEP, this is especially needed for the measurements
 of \ft at large \qsq and for the determination of \ftc.
 When the differences between the H1 and ZEUS results on di-jet production
 are resolved, these data will be a valuable input to further constrain 
 the partonic content of the photon.
%
%
\section*{Acknowledgments}
 I like to thank all those who contributed to the structure function session. 
 I greatly appreciate the help that I have been given by the speakers in
 preparing the summary talk.
%
%

\end{document}